\RequirePackage{fix-cm}
\documentclass[10pt]{article}  
\usepackage{graphicx}
\usepackage[margin=1.2in]{geometry}
\usepackage{amsmath}
\usepackage{booktabs} 
\usepackage{flushend}
\usepackage[ruled]{algorithm2e} 
\usepackage{caption}
\usepackage{varwidth}
\usepackage{wrapfig}
\usepackage[export]{adjustbox}
\usepackage{color}
\usepackage{authblk}

\newcommand{\para}[1]{{\vspace{4pt} \bf \noindent #1 \hspace{10pt}}}

\begin{document}

\title{Gender Differences in Participation and Reward on Stack Overflow
}


\author{Anna May$^{\rm\mathsection}$\hskip0.5cm Johannes Wachs\thanks{Corresponding author: johanneswachs@gmail.com}\protect\phantom{\footnotesize 1}$^{\rm\dagger}$\hskip0.5cm Aniko Hannak$^{\rm\ddagger}$ 
\\
\small{$^{\rm\mathsection}$Department of Economics and Business, Central European University\\$^{\rm \dagger}$Department of Network and Data Science, Central European University\\ \small{$^{\rm\ddagger}$Vienna University of Economics and Business \& Complexity Science Hub}}}
\date{}

\maketitle

\begin{abstract}
Programming is a valuable skill in the labor market, making the underrepresentation of women in computing an increasingly important issue. Online question and answer platforms serve a dual purpose in this field: they form a body of knowledge useful as a reference and learning tool, and they provide opportunities for individuals to demonstrate credible, verifiable expertise. Issues, such as male-oriented site design or overrepresentation of men among the site's elite may therefore compound the issue of women's underrepresentation in IT. In this paper we audit the differences in behavior and outcomes between men and women on Stack Overflow, the most popular of these Q\&A sites. We observe significant differences in how men and women participate in the platform and how successful they are. For example, the average woman has roughly half of the reputation points, the primary measure of success on the site, of the average man. Using an Oaxaca-Blinder decomposition, an econometric technique commonly applied to analyze differences in wages between groups, we find that most of the gap in success between men and women can be explained by differences in their activity on the site and differences in how these activities are rewarded. Specifically, 1) men give more answers than women and 2) are rewarded more for their answers on average, even when controlling for possible confounders such as tenure or buy-in to the site. Women ask more questions and gain more reward per question. We conclude with a hypothetical redesign of the site's scoring system based on these behavioral differences, cutting the reputation gap in half.
\end{abstract}

\section{Introduction}

As coding skills find their way into the basic requirements of many well paying jobs~\cite{burningglass-2016-coding}, the underrepresentation of women in technical fields is becoming an increasingly salient issue~\cite{techrep-2014-womentech}. Recent efforts to reduce this discrepancy are multi-faceted: while communities aimed at teaching girls or women to code focus on issues related to self-confidence and gender stereotypes~\cite{RePEc:mtp:titles:0262033453,ahuja2002women}, more IT companies and schools are promoting diversity and fighting discrimination~\cite{clayton2002ten}. Online resources also provide significant and informal opportunities for people who want to learn how to code, from free courses to entire communities for learning, discussing, and collaborating~\cite{burningglass-2016-coding,online-classes-wired,lerner-2002-open}. Two prime examples of the latter are Stack Overflow and GitHub. One would hope that the digital nature of these new ``knowledge marketplaces'' could democratize knowledge and help to mitigate existing inequalities. Yet, research finds that the opposite has happened. Contribution rates for women in open-source programming communities such as GitHub or Stack Overflow are even lower than female's presence in the IT labor market.\footnote{7.6\% of the users participating in the 2017 Stack Overflow survey (https://insights.stackoverflow.com/survey/2017) identify themselves as women, while in 2015, 25\% of the computing jobs were held by women in the U.S. ~\cite{womentech} } These trends align with findings from studies on other open-source communities and knowledge creation platforms, including Wikipedia and OpenStreetMap~\cite{stephens2013gender,wiki-16-silent,git-harrass}. These studies formulate a variety of hypotheses to explain this effect, including the impact of gender roles and stereotypes, lower confidence and risk aversion among women, and the asymmetric threat of harrassment.

In this study we explore reasons behind low participation and success rates of women on Stack Overflow, the largest Q\&A platform for programming and an important resource in the open source IT world. Over time, Stack Overflow has grown into a large database of knowledge which people use at all stages of learning how to program. Questions vary in difficulty and specificity, and the coverage of topics evolves essentially in sync with coding itself. Beyond being a knowledge base, the site also serves as a social platform, job search site, and recruiting tool - it is an important hub in the IT ecosystem. 

We collect a gender-balanced sample of over 20,000 user profiles, and use them to investigate the differences in the participation and success of men and women on the site. We frame our analysis in terms of the following questions. Do men and women have different levels of success on Stack Overflow? If so, is it because of differences in how they participate on the platform? We find that the answer to both questions is yes, and follow up by probing the differences in rewards for different kinds of participation on the platform.

More specifically, we find significant gender gaps in activity: women are more likely to ask questions, while men provide more answers and cast more votes. Votes are positive or negative evaluations of other user's questions and answers. Users gain and lose reputation points, the primary measure of success on the platform, for receiving up and down votes. We also observe that men are significantly more successful on the site, measured by their collection of reputation points. Using the Oaxaca-Blinder decomposition ~\cite{oaxaca1973male}, a method from economics that to the best of our knowledge has not yet been previously applied to measure gender disparities in online communities, we decompose the outcome differences between men and women in terms of differences in their activity. Our models show that question and answer behaviors and other user- and community-level features explain a large portion of the success gap. 11\% of the reputation gap remains unexplained.

In the final part of the paper, we explore the consequences of a hypothetical redesign of the site's reward system. The proposed alternative scoring system equalizes the rewards for well-liked questions and answers, a simple and justifiable change which does not penalize any group of users in absolute terms. We find that the median woman is marginally more successful than the median man under this revised success measure, reversing the situation under the current system. However even with our recommendation men are still significantly more successful on average due to their overrepresentation among the top users. The recommendation may alter site dynamics as users will be incentivized to ask more and better questions. Given Stack Overflow's stated aim to build a universal knowledge base, we believe that such a shift in the dynamics is in line with the spirit and goals of the platform.

In general however, our findings suggest that fundamental remedies may be needed in order to encourage women to participate more and in different ways. Given the increasing importance of Stack Overflow and similar sites in both the labor market and knowledge creation, our findings underscore the importance of design decisions and interventions even in well-intentioned and organically grown online communities.
\section{Related Work}

In this section we outline recent work on gender gaps in IT and on the web. We also survey studies which investigate online platforms to detect and measure inequalities.

\paragraph{Gender Gaps and IT} In the US a woman earns about 80 cents for every dollar a man earns. Even though the gap has been shrinking since 1960, it is still present at most educational levels and lines of work~\cite{blau2016gender}. The gap is larger within traditionally male-dominated fields such as computing. How do these fields remain male-dominated? Research shows that women are significantly underrepresented in academic fields ``believed to require attributes such as brilliance and genius''~\cite{leslie2015expectations} including computer science. When they do choose to enter these fields, they have higher drop-out rates~\cite{jadidi2017gender} and have a harder time being successful because of ``masculine'' culture, discrimination, or the handicap of lower self-confidence~\cite{bentley2003gender}. Computer science is one of the fields where gender-based occupational segregation is still strong. While 57\% of all employees in the US are women, only 25\% of the employees in computing are women. They earn only 18\% of the bachelor's degrees in computational sciences~\cite{lehman2016women}. Between 1980 and 2010, 88\% of all the information technology patents were introduced by male-only teams, which shows that the technology we use is invented by a strongly male-dominated community~\cite{womentech}. This may worsen the situation, as studies show that men have an advantage over women when using tools designed by other men~\cite{beckwith2004gender,beckwith2006tinkering}.

\paragraph{Measuring Inequalities} Studies investigating gender and racial inequalities in online communities and labor markets find that the gaps are just as prevalent and relevant online as offline~\cite{wachs2017men,hannak-2017-cscw,thebault-2015-cscw,NBERw22776}. Many of these studies are concerned with discrimination based on information available on user profiles~\cite{terrell2017gender}, social feedback by the community as a manifestation of offline discrimination in online context and algorithms reinforcing existing gender biases~\cite{sweeney-2013-google,sandvig2014auditing,marom2014gender}. Scholars are also drawing attention to the legal aspects of discrimination and labor market protections in the online world~\cite{barzilay2016platform}. 

An online platform does not need to have a financial purpose to create or reinforce offline gaps in participation and success. Several authors have studied open-source communities and their inequalities. Previous research shows that the most frequently used online knowledge sources are often created by a small minority, because cultural and algorithmic features of the platform discourage women or other underprivileged groups from contributing and editing. Studies on Wikipedia find that women are underrepresented as editors and also as subjects of the content leading to a skewed representation of knowledge~\cite{lam-2011-wikisym,reagle-2011-ijc,wagner2016women,menking2015heart}. Similar patterns were found on OpenStreetMaps and Google MapMaker where the features identified on digital maps catered to men's tastes, as men contribute more than women~\cite{stephens2013gender}. The underrepresentation of women is more pronounced in content creation than participation. A recent study of Wikipedia~\cite{shaw2018pipeline} finds evidence of a leaky pipeline: while women are not significantly less likely to have heard of Wikipedia or visited the site, they are significantly less likely to know that the site can be edited or to have made a contribution.

Stack Overflow is itself a well-studied platform. Vasilescu et al. show that women are underrepresented in this community~\cite{vasilescu2013stackoverflow}. Interviews with a sample of Stack Overflow users highlight the barriers women have to greater participation. Women respondents listed the lack of awareness of some site features, the intimidating community size and their fear of lacking adequate qualifications as main barriers to participation~\cite{ford2016paradise}. Recent work by Ford, Harkins, and Parnin highlights an important effect of the gender imbalance on user activity: women are more likely to engage with a post on Stack Overflow if they see other women in the conversation~\cite{ford2017someone}. This finding is both promising, highlighting a potential virtuous cycle of increased engagement by women, and worrisome, as higher turnover among women could compound.

Users on the site can collect badges, tokens awarded to users for specific actions and activity, and past work has shown that this steers and influences user behavior~\cite{anderson2012discovering}. However, research shows that gamification does not impact women's and men's behavior in the same way. The disparate influence of gamification on men and women has been observed in educational settings, in workplace and on online platforms as well~\cite{herzig2015workplace,pedro2015does}. In an elementary school environment, a gamified educational virtual software supporting math teaching significantly improved boys' learning motivation, but it had no effect on girls' motivation or performance~\cite{pedro2015does}. One potential explanation, backed by experimental evidence, is that men are socialized to have a greater preference for competition than women~\cite{niederle2007women}. 

Methodologically, economists have a long history of estimating gender disparities. The Oaxaca-Blinder decomposition ~\cite{oaxaca1973male,blinder1973wage} is a widely used econometric tool to disentangle the reasons behind gender differences in various outcome variables. It has also been used to study causes of obesity in different racial groups~\cite{sen2014using}, differences in career advancement prospects of men and women~\cite{chen2010examining}, and the difference in labor market outcomes between agency-endorsed and independent job-seekers~\cite{stanton2015landing}. Education researchers have also used Oaxaca-Blinder to explore differences in why women and men aspire to major in computer science, and how this changes over time~\cite{sax2017anatomy}.

\section{Background on Stack Overflow}

In this section we describe Stack Overflow as a website and a community. Stack Overflow, founded in 2008, is the largest Q\&A site for computer programming. Today, the site hosts over 16 million questions and 24 million answers, and it has a global Alexa rank of 63~\cite{Alexa-SO}. Previous work on Stack Overflow has highlighted its importance to the programming community as a hub of knowledge-sharing~\cite{vasilescu2014social}. According to creators, their goal was to design a free access platform serving users with a high quality knowledge base~\cite{joel-blog}. Indeed, today programmers of any level or type turn to Stack Overflow as part of their daily routine and the site is usually among the top results in Google searches for programming related queries. In this way, the knowledge shared on Stack Overflow is reused beyond the initial exchange between question-asker and answer-giver. Its user-base also significantly overlaps with that of popular code repositories such as Github~\cite{vasilescu2013stackoverflow}.

Stack Overflow also has influence on hiring/recruiting in the IT sector. Users can search for employment opportunities on the site's job boards. Moreover, Stack Overflow provides opportunities for them to demonstrate credible, verifiable expertise. Indeed, IT companies and recruiters often look for Stack Overflow profiles when trying to fill positions~\cite{IT-sourcing}. This is facilitated by a recently developed resume service on the site: users can turn their profile and activity into a standardized, searchable resume, ready for inclusion in the site's database of jobseekers. In this way Stack Overflow as a platform is a becoming a significant labor market matching service.

Stack Overflow's knowledge base (namely the questions and answers that have been posted) are freely available to the public - no registration required. In order to create content, however, users have to sign up using an email address or social media account. Every account has an associated profile page which tracks a user's activity history and accomplishments on the site. Users can also enhance their profiles with biographical information, contact information, and an image. The large disparity between the number of unique monthly visitors (estimated by Quantcast to be 50 million in March 2018) and the number of active accounts made in the history of the site (less than 10 million as of March 2018, with far fewer active accounts), indicates that the vast majority of the site's users are passive: using the site's knowledge without making contributions of their own.

Registered members can post questions and answers, vote on or edit the questions and answers of other users, and interact with posts using comments. The up and down voting functionality serves as a natural content moderation, users can boost useful questions and answers and subsequent visitors have an easier time finding them. There are also elected official moderators among the community members, who can delete, modify content, and merge repeated questions into one topic. 

Aside from the potential for its open and public-facing nature, Stack Overflow also has a gamification aspect~\cite{anderson2012discovering}. Specifically, participation leads to users earning various reputation points and badges. Reputation points are primarily received for upvoted questions and answers. Users receive some moderation privileges when they accumulate enough reputation points. Bronze, silver, and gold badges can be earned through a variety of activities, for example for receiving some number of upvotes on a question or answer, editing posts for clarity, or even for visiting the site for a number of consecutive days. Gamification is a common method used both to increase user engagement and to steer users towards specific behaviors deemed to benefit the community. Indeed, previous studies show that the badge system of SO has a motivating effect on the community. However, as past research indicates that men may respond more to gamification than women, this may exacerbate gender inequalities in participation~\cite{pedro2015does,herzig2015workplace}. We note that a user's reputation and badge counts are immediately visible next to any question they ask or answer - offering a signal of the user's presence and participation on the site to others.

\section{Data and Methods}

\begin{table}[t]
	\centering
	\small
	\begin{tabular}{ll}
		\toprule
		\textbf{Outcome }& \textbf{Reputation} \\ \hline
		Answer Upvoted     & +10                  \\
		Answer Downvoted   & -2, (-1 to downvoter) 	\\
		Answer Accepted    & +15, (+2 to acceptor) \\
		Question Upvoted   & +5                    \\
		Question Downvoted & -2 \\
		Offer Bounty       & -Bounty Value          \\
		Answer Wins Bounty &+Bounty Value          \\
		Answer Marked Spam & -100                  \\
		Edit Accepted      & +2 (max 1000/user)   \\
		\bottomrule
	\end{tabular}
	\bigskip
	\caption{How users gain or lose reputation points on Stack Overflow.}
	\label{Reptable}
\end{table}

In this section, we present a more detailed overview on how the Stack Overflow website works and the data we gathered about users. We also give a thorough outline of the features we extracted or created and that will serve the basis of the upcoming analyses.  We now present our data collection and labeling methodology. Additionally, we introduce our dataset, focusing specifically on how the data breaks down along gender lines.

\subsection{The Website}
Users who create accounts on Stack Overflow can ask and answer questions as well as comment on questions or answers. For easy navigation between question and topics, users label questions with tags, indicating the topic of the question (for example if the question is about a specific programming language or algorithm). They gain reputation points, our fundamental measure of success, by receiving explicit positive feedback called ``upvotes''  on their questions or answers. We outline the ways users accumulate reputation in Table~\ref{Reptable}. Any user who accrues 15 reputation points gains the ability to upvote questions and answers\footnote{https://stackoverflow.com/help/privileges/vote-up} Stack Overflow also rewards specific behaviors with badges, which are tokens given for some kind of accomplishment (for example visiting the site every day for an extended period of time, receiving a set number of upvotes for a question they ask etc.).

We use the Stack Overflow API\footnote{https://api.stackexchange.com/docs} to collect information on all users with at least 100 reputation points, as these users can be considered active on the website (they are granted the basic rights to comment, upvote, flag and edit on Stack Overflow). In all, we collected data on 565,171 users. To supplement the information provided by the API, we scraped data on users' activity, including the badges they collected, the tags they used, and the count of questions and answers they posted. 

\subsection{Feature Creation}

Several of the features we use in our analysis can be extracted directly from user profiles. First, we note users' meta data, including their biography text, sign-up date, and whether they link to a personal website or social networking accounts such as Twitter, Linkedin, or Github. We operationalize these features in our models as a self-promotion index which takes a value between 0 and 1 depending on the proportion of self-promotion fields that the user has filled out. Within the biography field we check the text for the substrings ``senior'', ``lead'', ``head'', and ``manage'', assigning a dummy to each user taking the value 1 if they list any of these leadership or senior position indicators in their bio.

Second we quantify their activity on the site by how many questions they ask and answer, how often they edit posts, how many upvotes and downvotes they cast, and how often they make posts with which tags. Finally, we have information about user's outcomes and success on the site from their reputation scores and the number and types of badges they receive.

\paragraph{Gender Inference}

Inferring gender of individuals from their online profiles is a complex problem. We apply a two-step approach to infer  user gender, first using \textit{genderComputer}~\cite{VasilescuIWC13}, a tool specifically created to infer gender of Stack Overflow users from their given usernames and location. GenderComputer considers a variety of string manipulations (for example reversing ``Nohj'' to get ``John'') to . Location can provide additional accuracy by distinguishing, for example, between an Andrea from the UK (likely a woman) and one from Italy (likely a man). This method classified the users from our sample into 238,150 male, 24,717 female, and 302,304 unidentified users. In order to evaluate the quality of this classification, we manually examined 100 users classified as men and 100 classified as women. We found that while the method performed very well on men (97\% agreement with our manual check), our manual check agreed only in 44 out of 100 cases of women. This replicates the recent finding by Ford et al.~\cite{ford2017someone} that genderComputer sacrifices precision for greater recall when inferring women users.  

The second step of our inference seeks to correct this bias by applying a more conservative method based only on first names and location called \textit{Gender Guesser}. By considering only users rated as likely male or likely female by both methods, we are left with a smaller but more accurate sample. 10,571 users are rated as highly likely women by both methods. We randomly choose 10,571 likely men (again classified as such by both methods) to obtain a balanced sample. We repeated our manual check of a random sample of accounts finding 96\% agreement with our classification of men, and 84\% agreement with our classification of women. This ensemble approach resembles Ford et al.'s modification of genderComputer to focus on the detection of first names within the username~\cite{ford2017someone}.


We acknowledge several limitations and drawbacks to our approach to inference. First, we make the simplifying assumption that gender is binary. We argue that this is a fundamental limitation of examining questions about gender differences using harvested data. Second, discarding alias usernames builds on the assumption that men and women are equally likely to adopt user names that can be mapped to their respective gender, and that this mapping does not substantially impact our hypotheses. However, previous research has shown that anonymity impacts behavior~\cite{robertson-2017-estimates}, and it is possible that some users utilized an anonymous name in order to establish an independent identity. Such name selection is highlighted by the literature on gender swapping in online communities~\cite{bruckman1996gender,szell2013women}, where, for example, women may pose as men if they feel that they will be taken more seriously or to avoid harassment. We also note the limitations of the geographic component of our inference: a minority of users include location data, and a given location may not reflect a user's origin (for example if an Italian man named Andrea moved to the UK). Finally, name-gender databases have been shown to have significantly less accuracy when used to infer gender for non-European names~\cite{karimi-www16}.

Despite these limitations, we argue that our focus on identifiable names provides the best possible data to test our hypotheses of gender behavioral and outcome differences on Stack Overflow. By limiting our dataset to users where we are highly confident about our gender inference, we gain greater confidence in the estimates of our econometric models. Moreover, our analysis includes robustness checks with 5, 10, 20, and 50 percent of our gender labels in the balanced sample randomly shuffled. These test help us better understand the effect of potential classification errors on our results. See details in Section~\ref{subsec:reg}.

\paragraph{Detecting User Communities}

\begin{figure}[t]
	\centering
	\includegraphics[width=0.6\columnwidth]{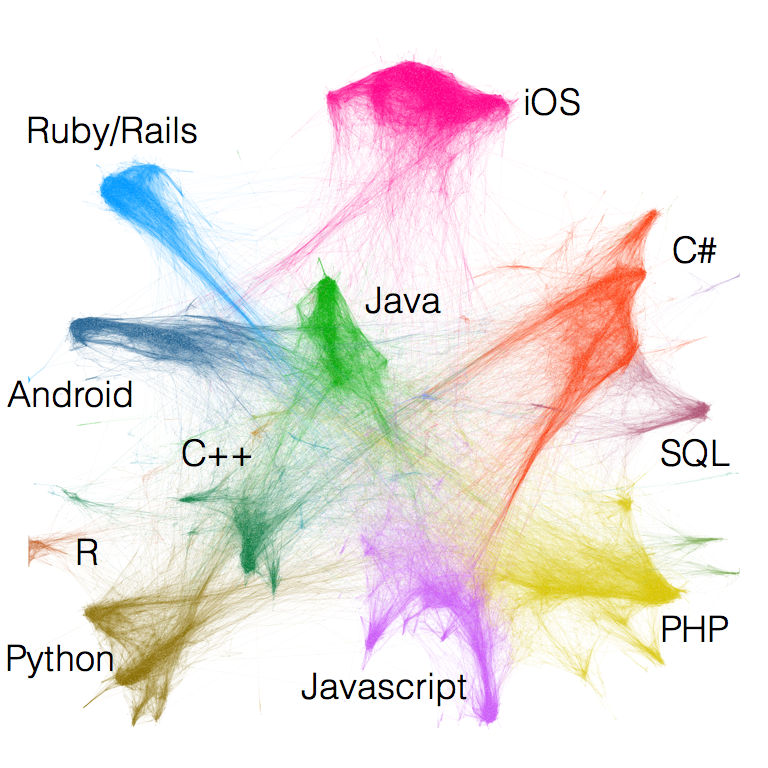}
	\caption{User-user tag similarity network. Two users are connected if they have statistically similar tag-use patterns. Colors indicate communities detected using a community detection algorithm. Labels correspond to the most frequently used tag in each community.}
	\label{network}
\end{figure}

Given the size of the site and the diversity of topics that its users discuss, we consider that coherent communities of users may exist with significantly different patterns of behavior, norms, and outcomes for men and women. For example, users active in a more diverse community may be less likely to leave the site~\cite{vasilescu2015gender}, while women encountering other women are more likely to engage in a thread~\cite{ford2017someone}. Using a similar approach to ~\cite{bosu2013building}, we grouped users in communities by building a network where two users are connected if their posts often share the same tags. Specifically, we created a similarity measure between users by calculating a weighted Jaccard similarity measure, defined as $$s(u,v)=
\dfrac{\sum_{ t \in T} \min (t_{u},t_{v})}
	 {\sum_{t \in T}  \max(t_{u},t_{v})}$$

\noindent where $T$ is the collection of all tags used at least 200 times, and $t_{u}$ denotes the number of times user $u$ made a post with tag $t$. We then filtered the edges using Serrano's disparity filter~\cite{serrano2009extracting}, which, for each node, checks the weights on all its adjacent links against the null hypothesis that they are uniformly distributed. Each observed weight then has a p-value. We filter edges using this p-value (p\textless.01). The resulting network has approximately 150,000 edges connecting the roughly 21,000 users. We use the Louvain algorithm~\cite{blondel2008fast} to detect communities in this network. We tune the method's resolution parameter to find larger communities to facilitate a qualitative understanding of the communities found.\footnote{The modularity score, a measure of the overall quality of a partition, does not significantly change when we tune the algorithm for this purpose.} We plot the network, visualized using a force-layout algorithm, in Figure~\ref{network}. The nodes are colored by community membership.

We manually checked the most commonly used tags in each community and found many clearly interpretable communities. The prominent programming languages and frameworks we identify in the largest communities coincide with those found in other analyses of programming language use, for instance on GitHub~\cite{celinska2017programming}.  We describe the 10 largest communities, accounting for 80\% of our users, in Table~\ref{communitystats}. Note that we sampled the males to achieve a 50-50 male-female ratio in our dataset. We see small, occasionally statistically significant gender differences. We find that the C\#/asp.net, a Microsoft-developed software framework,  and Ruby/Rails, a web development framework, communities have the highest representation of men, while Android, a programming language for mobile phone applications, has more women. We find that Ruby/Rails is the community with the lowest incidence of downvoting. 

\begin{table}
	\centering
	\resizebox{\columnwidth}{!}{%
	\begin{tabular}{lllllll}
		\toprule
		\textbf{Description} & \textbf{\# of Users} & \textbf{\% Male} & \textbf{\% Downvotes} & \textbf{\% Rep. Last Year} \\ \hline		
		C\#/asp.net   & 2900     & 54\% & 5.5\% &    6\%  \\
		Java   & 2605     & 49\% & 5.6\% &    7.2\%  \\
		PHP   & 1941     & 51\% & 7.4\% &   6.5\% \\
		Android   & 1856     & 45\% & 5.5\% &     8.3\%  \\
		Python   & 1665     & 49\% & 5.1\% &   9.3\%  \\
		iOS   & 1548     & 49\% & 5\% &    6.9\%  \\
		Javascript   & 1526  & 48\% & 7\% &    8.9\%  \\
		C++   & 1390  & 51\% & 5.8\% &   5.9\% \\
		Angular/Node   & 886  & 51\% & 5.3\% &    14.3\%  \\
		Ruby/Rails   & 873  & 55\% & 4.2\% &   6.8\% \\
		\bottomrule
	\end{tabular}}
	\bigskip
	\caption{Descriptive statistics of the 10 largest user communities based on a network of tag-use similarity among our balanced sample of users. We describe each user community by interpreting the most frequently used tags in posts by its members. The last column refers to the share of the community's total reputation gained in the last year.}
	\label{communitystats}
\end{table}

As past work indicates, community structure has a significant impact on user behavior and the possibilities for gaining reputation~\cite{bosu2013building}. For instance, it may be easier to ask a new question or post answers in a newer community, for example on Angular/Node related questions, than in a long established community such as on C++. Therefore subsequent models explaining gender differences (see Results section) include fixed effects for user community. We also control for the size of the community, the percentage of the community that is male, and the percent of reputation generated by users in the community in the last year as a proxy for how new the community is.

\section{Results}
\label{sec:res}

\subsection{Descriptive Statistics: Men vs Women}

\begin{figure}[!t]
  \begin{minipage}[t]{0.45\textwidth}
    \centering
	\vspace{0pt}
      	\begin{tabular}{lll}
		\toprule
		\textbf{Mean Activity} & \textbf{Women} & \textbf{Men} \\ \hline
		\# Answers   & 19.5     & 39.8    \\
		\# Questions & 16.4   & 13.5 \\
		\# Edits         & 9.0    & 10.0 \\
		\# Upvotes    & 115.0 &    170.0    \\
		\# Downvotes & 14.6 & 21.9\\
		Account Age (days) &1718 & 1925  \\
		\bottomrule
	\end{tabular}
	\vskip 0.875cm
	\captionof{table}{Average activity levels across gender. Men answer 53\% more questions on average than women do, while women ask 18\% more.}
	\label{table:activitydiffs}
  \end{minipage}
  \hfill
    \begin{minipage}[t]{0.5\textwidth}
   \centering
   \vspace{-5pt}
   \includegraphics[width=\textwidth,valign=t]{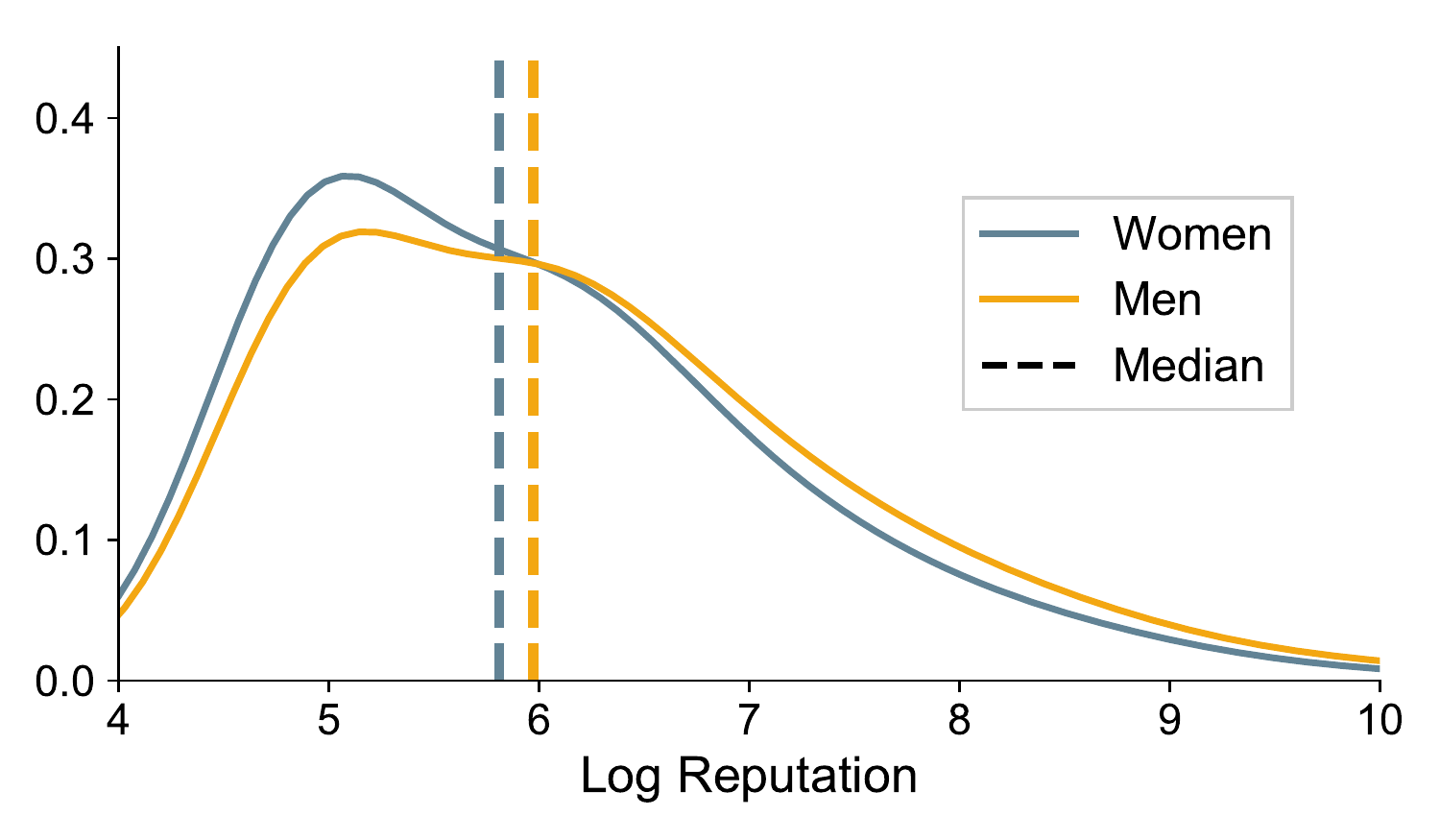}
   \vskip -0.2cm
   \caption{Kernel density estimates of the logged reputation scores of men and women.}
   \label{fig:Reputation}
   \end{minipage}
     \vspace{-10pt}
  \end{figure}

The first question we ask is whether we can see a difference in the outcome measures of men and women on the site.  Our key dependent variable is reputation, and we see that there are significant differences between men and women. The average reputation score is 1703 for men and 942 for women. In other words, women have on average 55\% of the reputation of men. The median woman has 73\% of the reputation of the median man, suggesting that many of the top reputation earners are men. The log-transformed reputation group averages are 6.1 for males and 5.8 for females, corresponding to the geometric means of 461 and 332, respectively. All differences are statistically significant (using a Mann-Whitney U test, p \textless .001). We plot the densities of the log reputation scores of men and women in Figure~\ref{fig:Reputation}.

We also note differences in average activity levels, outlined in Table~\ref{table:activitydiffs}. In contrast to a 2012 study of men and women on Stack Overflow~\cite{vasilescu2012gender}, which found that men are more active on the site across all measures, we find that women are more likely to ask questions. There are several possible explanations for this finding, for instance that the patterns of behavior on the site have changed, or because of differences in our approach to gender inference (i.e. having a lower false-positive rate among our likely women) or data selection (i.e. considering only users with at least 100 reputation). Indeed follow up work by the same authors of the 2012 study find that when controlling for overall length of engagement, women ask more questions~\cite{vasilescu2013gender}.

\subsection{Analysis} 
As outlined in the previous section we find several differences in both activity and outcome between men and women. How does the former impact the latter? We introduce a series of possible explanations for this difference, and check to see if controlling for these confounds in a regression framework can reduce or eliminate the gender gap. 

\begin{figure}[t]
\centering
    \includegraphics[width=0.9\textwidth]{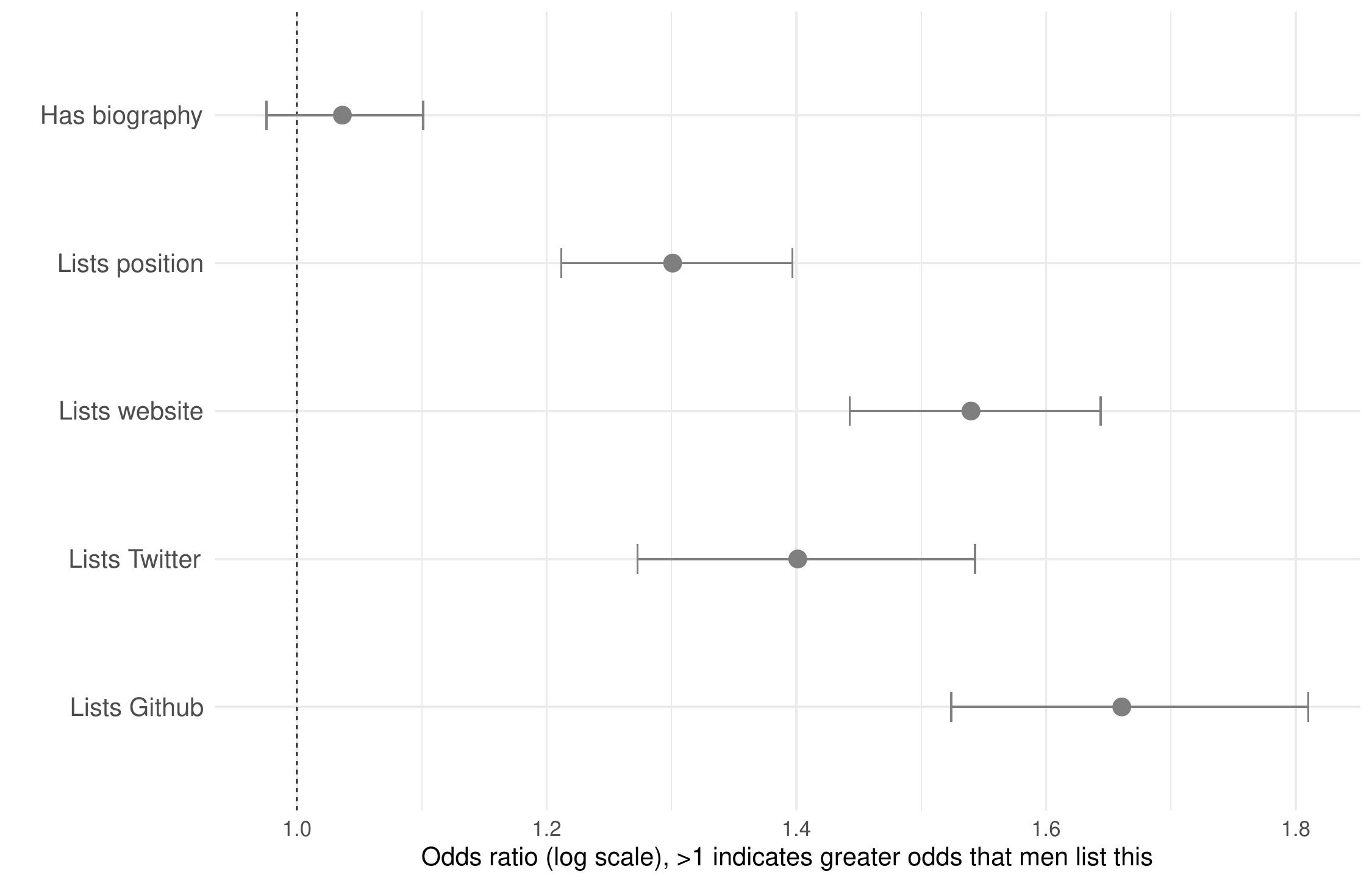}
    \caption{Differences in self-promotion using a propensity-score matched sample of men and women. Men are significantly more likely to fill in the position, website, Twitter and Github fields. Whiskers mark 95\% confidence intervals.}
    \label{selfpromotion}
\end{figure}

First we examine the differences between how men and women share information about themselves on the site. Similar to previous findings on LinkedIn~\cite{ICWSM1715615}, we find that men are significantly more likely to fill out their biography, to link to their Github, LinkedIn, Twitter, or personal websites on Stack Overflow. We plot the matched log-odds ratio in Figure~\ref{selfpromotion}. Second, we consider differences in tenure and find that on average men have been on the site significantly longer than women. Third, we consider that women may be overrepresented in certain communities with different norms and behaviors. We find limited evidence for gender segregation across communities, but nevertheless include community fixed-effects in later modeling efforts. Fourth, we examine differences in activity. We find that men are more likely to answer questions, while women are more likely to ask questions, with both differences significant according to Mann-Whitney U tests (p \textless 0.001).

\paragraph{Regressions}
\label{subsec:reg}
We use a linear regression framework to explore gender differences in user reputation. We examine the relationship between activity measures such as number of questions asked, number of answers given, and number of votes cast, and reputation (log-transformed), while controlling for potential confounders such as tenure. We report our findings in Table~\ref{interactiontable}. In a simple model controlling only for tenure, we find that users posting 1\% more answers are expected to have 0.50\% more reputation. On the contrary, the impact of asking an additional questions is an order of magnitude smaller. This is an inherent feature of the current scoring system, see Table~\ref{Reptable}. The coefficient on the male term is positive and significant but close to 0.

Next we include controls for the users's propensity to disclose information in their profile and community measures, such as the share of men in the community and the share of total reputation earned by members in the last year. This latter measure proxies the age of the community or the newness of its topics. We find similar coefficients as in the previous model.

\begin{figure}
    \includegraphics[width=\textwidth]{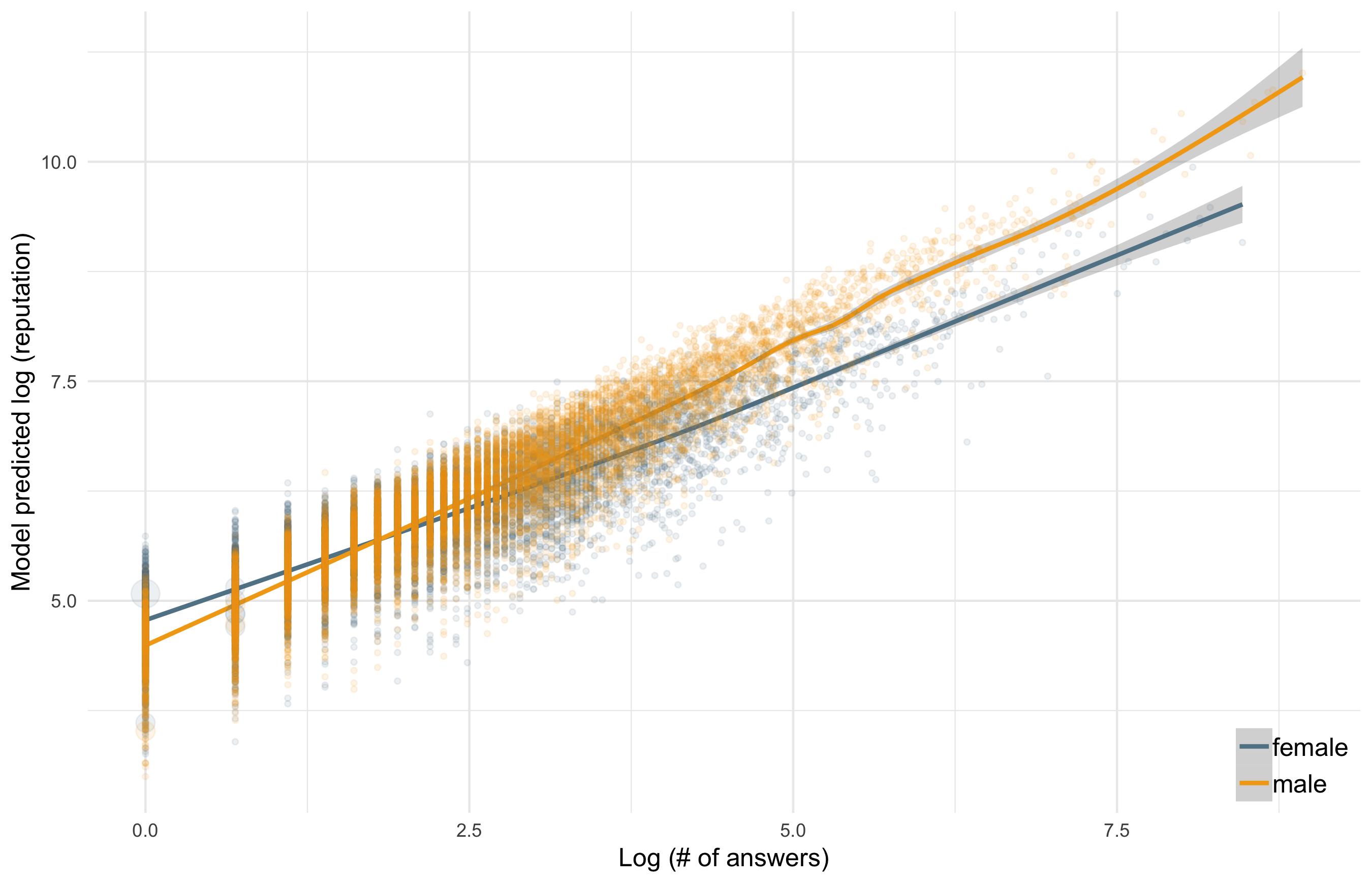}
    \caption{We plot the marginal effect of the number of answers a user posts on his or her reputation by gender according to Model 3 in Table~\ref{interactiontable}. The model controls for tenure, activity, self-promotion indicators, and community-level features. Note the difference in slopes between the genders, indicating that men get more reputation points for additional answers.}
    \label{interactionplot}
\end{figure}

Finally, we include interactions of the male term and both question and answer activity. The significant positive term on the interaction of gender and the number of answers posted shows that men gain more reputation for an additional answer than women. We visualize the interaction model in Figure~\ref{interactionplot}. We use an F-test to test the null hypothesis that the coefficient on the male term and its interactions in the third model are simultaneously equal to 0. The value of the F-statistic is 128 with 1 and 21119 degrees of freedom, and the test returns p-value less than $10^{-15}$. We reject the null hypothesis that the regression coefficient vector is the same for both genders. The significance of the interaction term justifies the use of a decomposition method in the following section to weigh the contributions of the various features to the overall outcome gap.

\begin{table*}[t]
\centering
\begin{tabular}{lccc} 
\\[-1.8ex]\hline 
\hline \\[-1.8ex] 
 & \multicolumn{3}{c}{\textit{Dependent variable: $\ln(\mathrm{Reputation})$}} \\ 
\cline{2-4} 
\\[-1.8ex] & (1) & (2) & (3)\\ 
\hline \\[-1.8ex] 
 Male & 0.04$^{***}$ & 0.03$^{***}$ & $-$0.25$^{***}$ \\ 
  & (0.01) & (0.01) & (0.02) \\  
 Answers Posted (log) & 0.50$^{***}$ & 0.52$^{***}$ & 0.45$^{***}$ \\ 
  & (0.005) & (0.005) & (0.01) \\  
 Questions Posted (log) & 0.08$^{***}$ & 0.09$^{***}$ & 0.09$^{***}$ \\ 
  & (0.004) & (0.004) & (0.01) \\  
 Votes Casted & 0.09$^{***}$ & 0.08$^{***}$ & 0.08$^{***}$ \\ 
  & (0.004) & (0.004) & (0.004) \\  
 Account Age & 0.49$^{***}$ & 0.50$^{***}$ & 0.50$^{***}$ \\ 
  & (0.01) & (0.01) & (0.01) \\  
 Male $\times$ Answers Posted &  &  & 0.13$^{***}$ \\ 
  &  &  & (0.01) \\  
 Male $\times$ Questions Posted &  &  & $-$0.01 \\ 
  &  &  & (0.01) \\  
 Constant & 0.79$^{***}$ & 1.35$^{***}$ & 1.43$^{***}$ \\ 
  & (0.07) & (0.38) & (0.38) \\  
\hline \\[-1.8ex] 
Observations & 21,142 & 21,142 & 21,142 \\ 
Self-promotion Controls & & YES & YES \\ 
Community Controls & & YES & YES \\ 
Adjusted R$^{2}$ & 0.61 & 0.62 & 0.63 \\ 
Residual Std. Error & 0.74 & 0.74 & 0.73  \\ 
F Statistic & 6,749.74$^{***}$  & 1,734.25$^{***}$  & 1,617.60$^{***}$  \\ 
\hline 
\hline \\[-1.8ex] 
\textit{Note:}  & \multicolumn{3}{r}{$^{*}$p$<$0.1; $^{**}$p$<$0.05; $^{***}$p$<$0.01} \\ 
\end{tabular}
\bigskip
 \caption{User reputation regressed on gender, user-level activity measures, controlling for self-promotion indicators and community-level features. The gender indicator interacted with the activity measures in the third model shows that men get more reward for posting additional answers than women do.}
 \label{interactiontable} 
\end{table*} 

We run two robustness tests on our results. First we randomly shuffle the inferred gender on subsets of our data to test the effect of error in our classification on the observed effect. We find that the male and male $\times$ number of answers terms from Model 3 in Table~\ref{interactiontable} both remain positive and significant if we randomly shuffle 5, 10, 20, or 50 percent of the users' gender classification. For instance a 20\% randomization of the gender labels shrinks the effect of the male term to -0.13 (from -0.25) and the interaction term to 0.11 (from 0.13), with both coefficients still significant at p$<$.01. The effect disappears when we completely randomize the gender labels.
 
Second, drawing on the observation that men are highly overrepresented among top users, we check our results dropping the top 1, 5, and 10 percent of users by reputation score. Our results are robust to this change. Finally, we combine the two robustness tests, randomly shuffling the gender label on 20 percent of our users, and dropping the top 1 percent. Both the male and interaction terms remain significant, albeit with smaller effect sizes (male coefficient: -0.07, interaction coefficient: +0.06, both significant at p$<$.01). Full model tables for the robustness tests are available on request.

\paragraph{Oaxaca-Blinder Decomposition}
The Oaxaca-Blinder decomposition is commonly used to measure and explain the causes of differences in averages between groups, including the wage gap between men and women~\cite{oaxaca1973male,blinder1973wage}. This decomposition for linear models allows us to observe the effect of differences in feature endowments that are used to predict the outcome between the groups (for instance that men may be on the site longer on average) and the effect of differences in how the features predict success - and thus different coefficients - between the groups (for instance if the same increase in tenure predicts a higher reputation boost for men than woman) separately. Neumark's elaboration (Equation~\ref{neum}) to the method introduced a tool to examine the effect of group endowments and coefficients compared to a vector of reference coefficients ($\beta^P$) computed from the pooled OLS regression~\cite{neumark1988employers}. The difference in the levels of explanatory variables weighted by the reference betas ($\triangle x \beta^P$) shows the part of the gap in the outcome variable which is explained by the differences in group averages (``explained''), while the explanatory variables weighted by the differences of the gender-specific and the pooled betas ($x^{male}(\beta^{male} - \beta^P) + x^{female}(\beta^P - \beta^{female})$) indicate the remaining, unexplained positive and negative biases. In the literature on wage differentials between groups, this second component is sometimes referred to as a measure of the discrimination present in a market. In our setting it captures the difference that would remain if women would have the same feature endowments as men.\footnote{For a longer exposition of the equations presented here see~\cite[p.~149-151]{o2008analyzing}.} 

\begin{equation} \label{neum}
\begin{split}
y^{\text{male}} -y^{\text{female}} =  \\
\triangle x \beta^P + \big[ x^{\text{male}}(\beta^{\text{male}} - \beta^P) + x^{\text{female}}(\beta^P - \beta^{\text{female}}) \big] 
\end{split}
\end{equation}

where

\begin{equation}
\triangle x = x^{\text{male}} -x^{\text{female}} 
\end{equation}

In our model estimating user reputation, we control for the potential explanations outlined above, including the number of answers given, questions asked, votes cast, and the age of the user account. We also include community-level features, and self-promotion dummies discussed before.

\paragraph{Twofold decomposition.}
According to the twofold Oaxaca decomposition (using the pooled OLS betas as reference coefficients following Neumark~\cite{neumark1988employers}), 89\% of the reputation differential can be explained by the effects of differences in the explanatory variables we used (number of questions and answers posted, number of votes cast, age of the account, average reputation change in the last year within the user's tag modularity class, self-promotion and leader dummies). The difference in the effect of number of answers posted online explains 75\% of the reputation differential, while the difference in the effect of account age provides an explanation for 22\% of it. The remaining (``unexplained'') 11\% of the reputation differential might be due to gender discrimination. Given that the inclusion of more features would likely decrease the difference explained by this component, we suggest that discrimination is a limited driver of reputation inequality on the site.

\begin{figure}[t]
  \includegraphics[width=\textwidth]{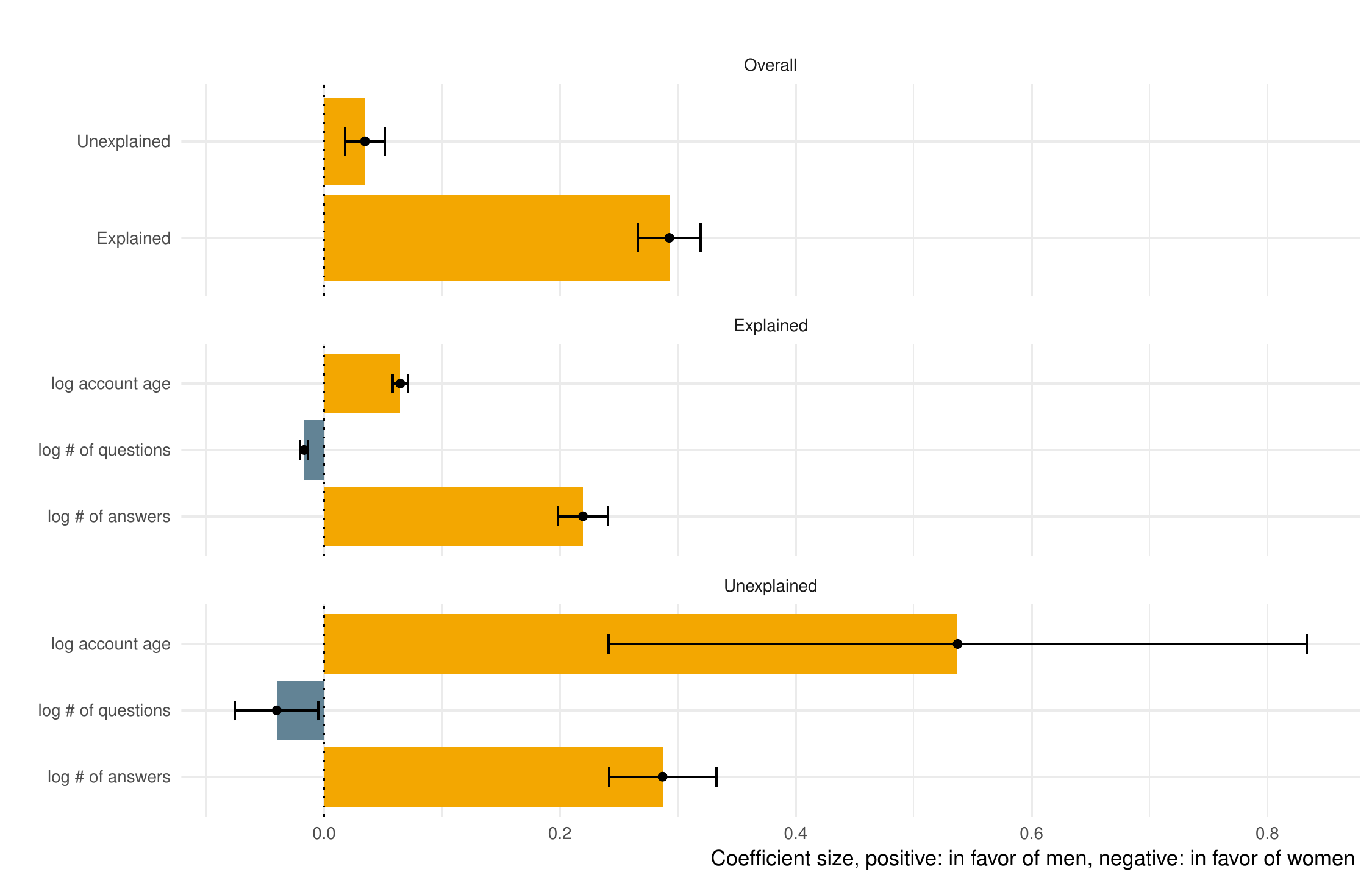}
	\caption{The Oaxaca-Blinder decomposition of the difference in log reputation between men and women. The \textit{Overall} subplot shows the difference decomposed into a part which is due to the known differences in endowments, accounting for 89\% of the difference, and unexplained effects accounting for 11\%. In the \textit{Explained} and \textit{Unexplained} subplots, each component is broken down into features. The difference in the amount of answers given by men and women as groups is by far the strongest explanatory factor. On all plots the whiskers indicate 95\% bootstrapped confidence intervals. In the latter two plots, only selected significant terms are shown.}
\end{figure}

\subsection{Discussion of the results.} We found that activity differences --- mostly the difference between the amount of answers given by men and women --- drive success inequality. There are a few theories in the literature than can explain the situation. Ford and her co-authors~\cite{ford2016paradise} found that women often hesitate to actively participate on the website because they fear they lack qualifications and because of the size and negativity of the community. While men, who are generally more competitive, thrive in this environment, women might be discouraged from answering due to these factors. We also note that the unexplained 11\% of the decomposition gap may indicate that women are treated differently on the site. 

We also see that women contribute differently to building the community's knowledge base: they are asking more questions. Stack Overflow's current system strongly incentivizes answering by rewarding upvotes on answers twice as much upvotes on questions. In the subsequent section we test how the outcome gaps would change if these rewards were equalized.

\section{Proposing an Alternative Reward System}
\label{sec:fix}

In this section we discuss one potential way to mitigate the gender differences in outcome and success on Stack Overflow: modifying the reward system. Our results suggest that differences in the rate at which men and women post answers account for the largest part of the reputation gap. On the other hand, women tend to ask more questions than men. As reputation points are almost entirely a function of the number of upvotes received on questions and answers posted, and because receiving an upvote on a question results in half of the reputation gain (+5) that receiving an upvote on an answer does (+10), we investigate what happens to the distribution of reputation scores of men and women if these rewards were equalized. In other words, we check if equalizing the rewards for good answers and good questions decreases the gender gap.

When Stack Overflow was launched in 2008, upvotes to questions and answers gave the receiver ten reputation points. In 2010 the rules were changed to their current format and reputation scores were retroactively altered for all users. In a blog post explaining the change, one of the co-founders of the platform cited three reasons for the decision to change the system\footnote{https://stackoverflow.blog/2010/03/19/important-reputation-rule-changes/}: 
\begin{itemize}
\item ``We know that answers have more intrinsic value than questions, and the reputation balance should reflect that.''
\item ``The question asker already enjoys a substantial benefit beyond reputation gain from upvotes on their question, namely, they get great answers to their question! Thus, the asker shouldn't need as much reputation gain.''
\item ``There are a few users who ask hundreds, sometimes even thousands of questions. Over time, these users generate a fairly sizable reputation entirely through the tiny trickle of upvotes gained by these questions. In a sense, we want to discourage question asking a little bit, and make sure that people who ask questions are doing it for the right reasons and not to generate reputation.''
\end{itemize}

Independent of the issue of gender disparities, we argue that the proposed change has merit when considered against these points, especially when considering the site's increased importance as a knowledge resource since 2010. We do not agree, for example, with the value judgement that answers have more intrinsic value than questions: without the question there would be no answer. Stack Overflow distinguishes itself from Wikipedia or textbooks as a knowledge resource by providing applicable answers to real-world user-generated questions. The service of asking a genuine question, the answer to which may seem simple to an expert, is part of what gives Stack Overflow its appeal above and beyond the example cookbook solutions available in many programming references or textbooks.We also note that a single question can generate multiple useful answers - suggesting that any one person answering a question can still learn from other answers. Finally, improvements in site moderation and semi-automated detection of duplicate questions~\cite{zhang2015multi} likely reduces the prevalence of reputation mining by asking repetitive questions.

Increasing rewards to good questions may help to make the site more inclusive by offering a less competitive and speed-oriented way to build one's reputation. For example, recent work on user strategies for gaining reputation on the platform focuses entirely on answers, finding that answering questions quickly (ideally first) is one of the best ways to quickly collect reputation~\cite{bosu2013building}. The authors also find that focusing on areas where there are fewer experts, or answering questions on off-peak hours are productive strategies. Such time pressures do not play the same role when one is asking a question.

\begin{figure}[t]
	\includegraphics[width=\columnwidth]{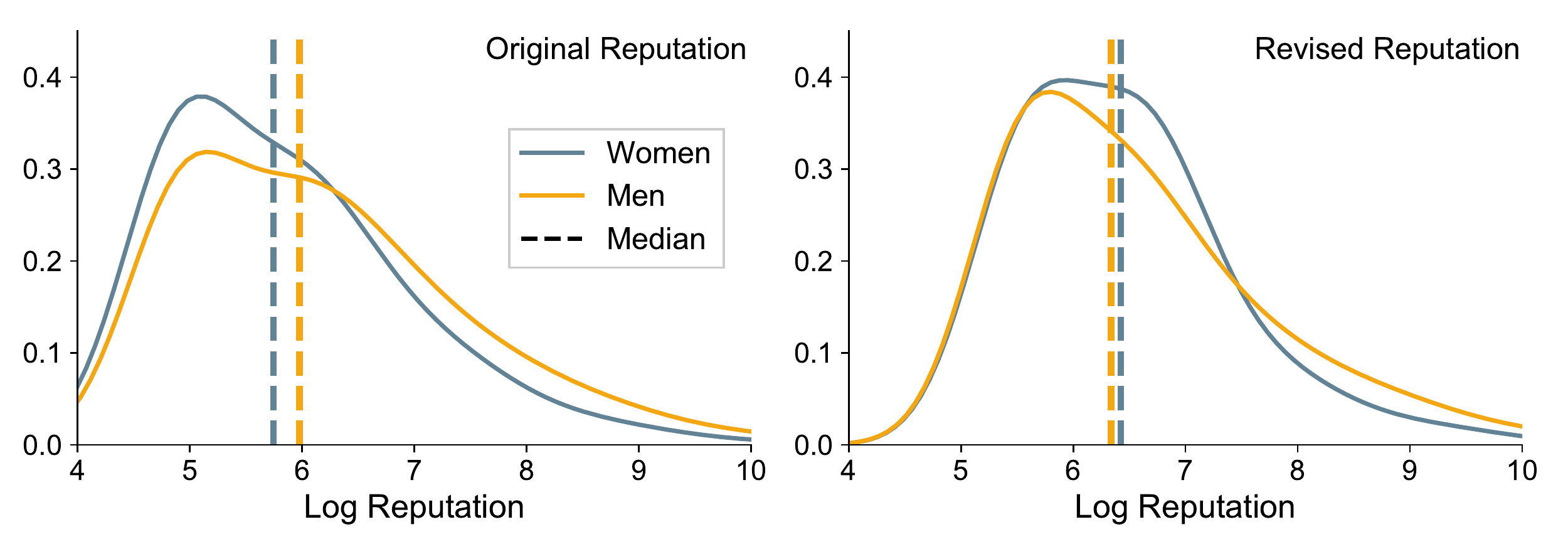}
	\caption{Distributions of (log) reputation and revised reputation for men and women. Dotted vertical lines indicate group medians. Note that although women have higher median in the revised reputation plot, men still have higher average reputation as evidenced by their overrepresentation in the long-tail of success.}
	\label{revisedrep}
\end{figure}


To calculate the revised reputation score, we collect additional data on each user's activity. We calculate the revised reputation scores of all users by counting question upvotes as being worth 10 points. Recall that women have on average 55\% of the reputation of men, and the median woman has 73\% of the reputation of the median man. Using the revised reputation, women have 71\% of the reputation of men on average, and the median female has \textbf{16\% more} reputation than the median male. We see this shift in Figure~\ref{revisedrep}, where we plot the distribution of log reputation and log revised reputation by gender. Indeed women have much lower variance: the low and high end of the distribution are proportionally much more male.

If additional reputation for asking questions even slightly increases the engagement of some women on Stack Overflow, peer effects might encourage other women to post~\cite{ford2017someone}. Yet we emphasize that this potential change is only one strategy to address gender disparities on Stack Overflow. Indeed, one should also be interested in why men are so much more likely to give answers, and how one might encourage women to give more answers. Furthermore, one must consider how all users would change their behavior if the scoring system stopped favoring answers. As men might respond more readily to gamification, it is possible that they would begin to ask more questions.

\begin{figure}
\centering
	\includegraphics[width=0.6\textwidth]{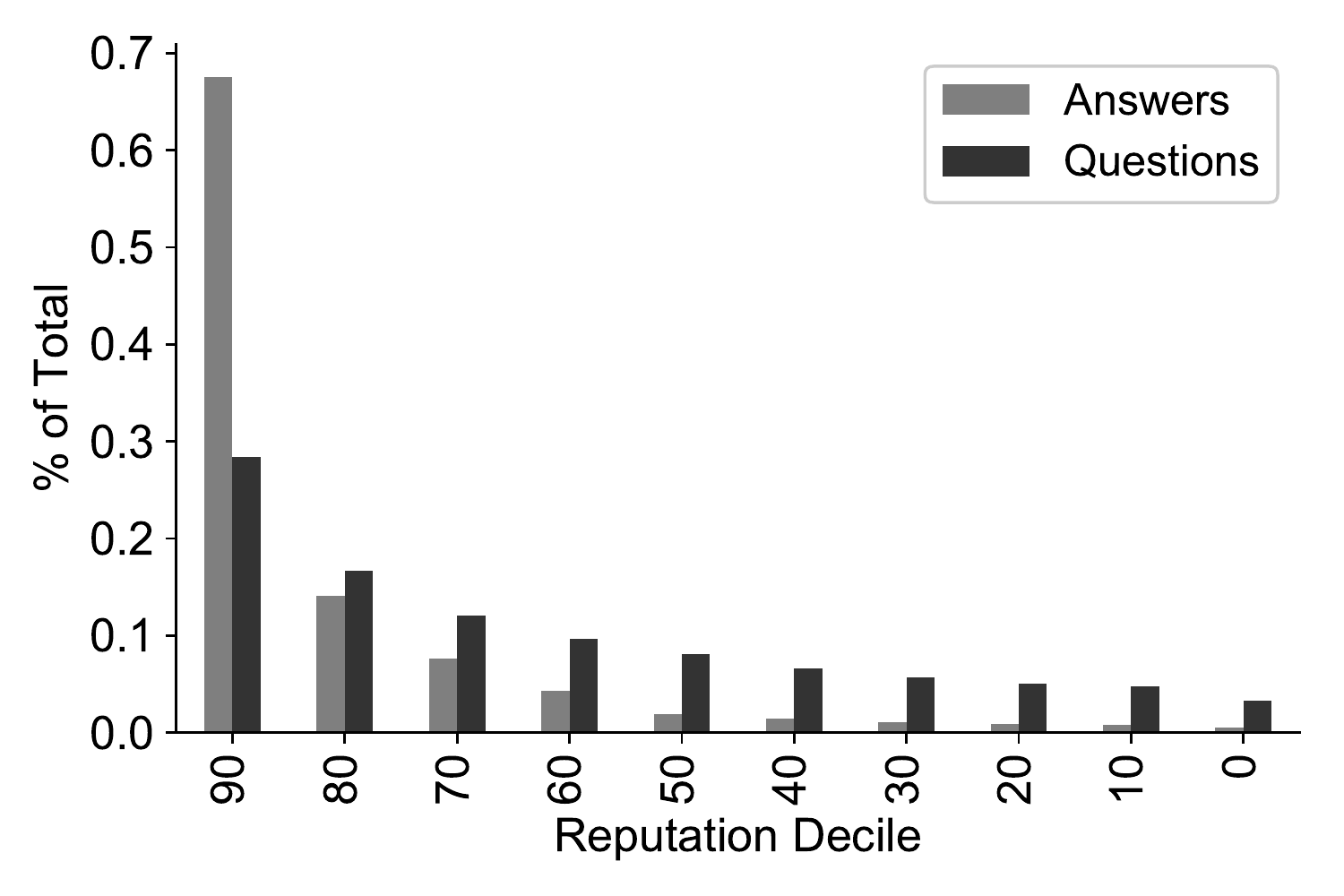}
	\caption{Share of total answers and questions posted by users group by reputation decile.}
	\label{aqdecile}
\end{figure}

In order to understand why the revised reputation scores do not change trends among the most successful users, we investigated how users of different reputation levels contribute to the site. Figure~\ref{aqdecile} shows the contribution rates broken down by activity type for each reputation decile. While questions are relatively evenly distributed across the groups, most answers are given by a small number of ``experts''. More precisely, 68\% of answers are given by the top decile of users, while only 28\% of questions are. While this result is somewhat intuitive given that the motivation for asking a question is a lack of knowledge, we think it brings attention to an important issue. Increasing rewards for questions encourages users from a broader spectrum of the population, most critically the learners who form an essential part of the site.

\section{Discussion}

In this paper we investigated gender differences in the activity and success of users on Stack Overflow. Women are highly underrepresented in the data we collected, not only compared to male users of the website but also compared to their presence in the IT labor market more broadly. Moreover, according to our success measures they are also less successful on the site than men. These differences can partly be explained by different participation or activity behavior. While women ask more questions than men, men are more active in all other forms of activity. Given that Stack Overflow's current reward system favors popular answers over popular questions, these differences in activity leads to large differences in outcome. Excluding this effect and controlling for important user and community level explanatory features, only 11\% of the reputation gap, which may be due to ``perceptional'' discrimination (among other confounders), remains unexplained. Given our finding that the current rewards system favors ``male behavior'', we propose a new reputation system which equalizes the reward for upvotes on questions and answers. We find that the new scoring system reduces differences in the group means, and in fact leaves the median woman with slightly higher reputation than the median male.

\para{Limitations}

Our analysis has several limitations. First, we emphasize that our data was selected from individuals who registered and managed to accumulate at least 100 reputation points. As research on gender inequalities has shown again and again, survival bias likely influences the features of women in our data. Hence we must acknowledge that the men and women in our data form a biased sample of all users of the site. The individuals who are active on the site are likely to be among those who are more able to take risks, more likely to prefer competition, and less vulnerable to harassment. This would lead to an underestimation of gender differences in behavior and outcome on the site. Moreover, a large part of Stack Overflow's audience is silent. Statistics from the site suggest that there are tens of millions of monthly visitors but only half a million accounts with 100 reputation points. It is likely that these roughly 500,000 users are less than 10\% of people who have used the site for learning. The true motivations for joining or staying silent is a very interesting question and could be addressed by gathering qualitative data~\cite{ford2016paradise}.

We also reiterate the issues we raised about inferring gender from online data. We may underestimate the participation of women by missing those who pose anonymously or as men. We also acknowledge a western bias in the method we used to detect gender: studies show that such methods suffer from significantly higher error rates on non-western names~\cite{karimi-www16}. 

Another drawback of our analysis is that we have no information on the quality of the questions and answers posted outside the user-level upvote and downvote scores. Since quality likely interacts with the audience's feedback and the existing reputation of users these are potentially important missing controls.

Lastly, we do not know the effect of current success on future success; in other words, whether the rich get richer. Research on other platforms shows evidence for such a reinforcement effect~\cite{muchnik2013social}. Our snapshot only allows us to hypothesize that such an effect may be present. More importantly, the limitations of our work highlight the need for a greater understanding of how success on Stack Overflow impacts success in the job market. We believe that these limitations can be addressed in future work by integrating experimental methods, user surveys, and the collection of longitudinal data about users' career paths on the site. 

\para{Impact}
We believe that we are at a crucial moment in the process of inclusion of women into the STEM and IT labor markets. One positive indicator of progress on the site is that women are more active question posters than men, while in a 2012 study men were significantly more active than women in all forms of activity~\cite{vasilescu2012gender}. This is perhaps a result of an increased focus of the Stack Overflow developers on improving the diversity and inclusiveness of their platforms~\cite{ford2017someone}, and is in line with changes observed in the annual Stack Overflow survey.\footnote{https://insights.stackoverflow.com/survey/2017} However it is not enough to encourage women to start learning, and it is important to continue efforts to support and include a diverse user population reflective of the broader population.

Though our findings and subsequent recommendation address a specific gender gap found on the site we analyze, we argue that there is more work to be done on why male and female behavior differs so significantly in the first place. The more the difference is due to the diversity of experience or perspective between the groups, the more our recommendation is a useful solution to gender gap in participation. Better recognition of the validity of alternative behaviors would likely encourage participation. On the other hand, given the literature on gendered barriers to participation in IT, we suspect that a large part of the behavioral difference is a legacy of these barriers and our recommendation is only a partial solution. In other words, if women are answering fewer questions because they have been discouraged from speaking up, fear harassment, or lack self-confidence, one cannot solve the overarching problem by increasing the rewards to questions. 

One promising recent development on Stack Overflow highlighting the need for more targeted intervention is the launch of a mentoring program for new users~\cite{ford2018we}. New users asking questions entered ``Help Rooms'' to obtain constructive criticism on their questions before posting on the main site. Mentored users asked higher quality questions and were more likely feel part of the community on Stack Overflow than control group users. In parallel with considering revisions to site design, more work is needed on how to scale this kind of intervention.

More broadly, we highlight the importance of auditing systems and evaluating the algorithms of open-source communities. Even though these systems have platforms that are more transparent, owners that are more benign, and financial impacts that are lower than large corporations like Google or Facebook, their social and labor market impacts can still be very large. As such, their owners share responsibility for the outcomes of women in the IT community, and the culture of open-learning more generally.

\section*{Acknowledgements}
We would like thank Agnes Horv\'at, Daniel Larremore, Piotr Sapiezynski, Kenny Joseph, and participants of Central European University's Gendered Creative Teams Workshop for helpful comments and insights. We thank Kristen Altenburger and Dorota Celinska for advice regarding matching and the Oaxaca-Blinder decomposition, respectively. We also acknowledge the comments of anonymous referees.

\bibliographystyle{spmpsci} 
\bibliography{biblio}

\end{document}